\documentclass[aps,prl,twocolumn,showpacs,preprintnumbers,amsmath,amssymb,groupedaddress]{revtex4}
\usepackage{tipa}
\usepackage{graphicx}
\usepackage{txfonts}

\begin{document}


\title{Polarizable vacuum analysis of electric and magnetic fields}


\author{Xing-Hao Ye}
\email[Electronic address: ]{yxhow@163.com}

\affiliation{Department of Applied Physics, Hangzhou Dianzi
University, Hangzhou 310018, China}


\date{\today}

\begin{abstract}
The electric and magnetic fields are investigated on the basis of
quantum vacuum. The analysis of the electromagnetic energy and force
indicates that an electric field is a polarized distribution of the
vacuum virtual dipoles, and that a magnetic field in vacuum is a
rearrangement of the vacuum polarization. It means that an
electromagnetic wave is a successional changing of the vacuum
polarization in space. Also, it is found that the average half
length of the virtual dipoles around an elementary charge is $a=2.8
\times 10^{-15}$m. The result leads to the step distribution of the
field energy around an electron, the relation between the fine
structure constant and the vacuum polarization distribution, and an
extremely high energy density of the electromagnetic field.
\end{abstract}


\maketitle


The quantum theory and experiments have approved that vacuum can be
polarized. The vacuum polarization can be used to interpret the
gravitation \cite{Dicke1957,Puthoff2002,Puthoff2005}, which reaches
an insightful description of the vacuum around the gravitational
matter with changeable permittivity $\varepsilon$ and permeability
$\mu$ \cite{Dicke1957,Puthoff2002,Puthoff2005} or graded refractive
index $n$
\cite{Dicke1957,Puthoff2002,Puthoff2005,Vlokh2004,Vlokh2005,
Vlokh2006,Vlokh2007,Ye2008-01,Ye2008-02}. A vacuum polarization
interpretation of the gravitational field endows the space of vacuum
with physical qualities. It is just what Einstein hoped and
predicted \cite{Einstein1920}. A reasonable extrapolation is that
the electromagnetic fields can also be interpreted as the effects of
vacuum polarization. Such an interpretation claims for a vacuum
whose properties are somewhat like those of dielectric medium
\cite{Lee1981}.

Exhilaratingly, facts and theories all indicate that vacuum is
actually a special kind of medium \cite{Ahmadi2006,Dupays2005}.
Casimir effect\cite{Emig2007,Lamoreaux1997} tells that vacuum is not
just a void, but a special physical existence full of zero-point
energy. Vacuum can be polarized by electromagnetic field, which
leads to the well-known effects of Lamb shift and anomalous magnetic
moment of the electron \cite{Ahmadi2006}. If the electromagnetic
field is extremely powerful, the vacuum will be excited to produce
$e^--e^+$ pairs \cite{Narozhny2004}. Dupays et al. pointed out that
the optical properties of quantum vacuum could be modified by
magnetic field, witch will influence the propagation of light
emitted by a magnetized neutron star \cite{Dupays2005}. Rikken and
Rizzo predicted that magnetoelectric birefringence will occur in
vacuum when magnetic and electric fields are perpendicularly applied
\cite{Rikken2003}. The phenomena that the propagation of light can
be modified by applying electromagnetic fields to the vacuum are
just similar to the Kerr electro-optic effect and the Faraday
magneto-optic effect in dielectric medium. This similarity between
the vacuum and the dielectric medium implies that vacuum must also
have its inner structure, which could be influenced by electric
charges or currents.

In this letter, both the electric field and magnetic field will be
analysed on the basis of vacuum polarization. The energy of electric
and magnetic fields will be figured out by using the intensity of
vacuum polarization. The electrical and magnetic forces will be
discussed considering the action of the nearby virtual dipoles in
vacuum. By doing this, the relation between the electromagnetic
fields and the quantum vacuum will be clarified. The electromagnetic
wave will then be described as an effect of successional changing of
vacuum polarization. Also, it will be shown that the virtual dipoles
around an elementary charge have a characteristic half length on
average, which leads to some interesting findings
further.\\

First, examining the electric displacement vector \textbf{D} in a
dielectric medium:
\begin{equation}
\mathbf{D}=\mathbf{P}+\varepsilon_0 \mathbf{E},
\end{equation}
where \textbf{P} represents the polarization of dielectric medium,
$\varepsilon_0$ is the permittivity of vacuum, \textbf{E} denotes
the electric field intensity in the medium. It is noticed that in
the equation $\varepsilon_0 \mathbf{E}$ can be interpreted as the
polarization of vacuum $\mathbf{P'}$ \cite{Dicke1957,Puthoff2002},
that is
\begin{equation}
\mathbf{P'}=\varepsilon_0 \mathbf{E}.
\end{equation}

The vacuum polarization is based on the knowledge that there are
virtual positive-negative charges, for example, but not limited to,
virtual $e^--e^+$ pairs, being randomly created and annihilated in
vacuum. Similar to a dielectric medium, the virtual charge pairs in
a vacuum will be polarized and aligned by the real charges, or by
the so-called ``external electric field''. For example, a vacuum
will be polarized in a spherical symmetry by a positive electric
charge $+Q$ as shown in Fig.~1, where the solid $\bigoplus$ denotes
the real charge $+Q$, the dashed $\ominus\oplus$ is a simplified
representation of a polarized virtual charge pair in the vacuum, and
the direction from the negative virtual charge to the positive one
indicates the direction of the electric field.

\begin{figure}
\includegraphics[width=2.6in]{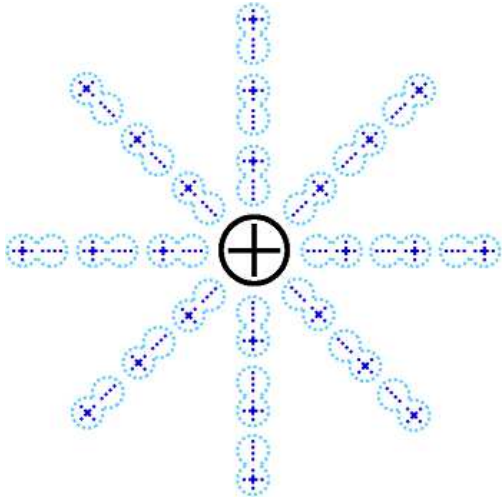} \caption{The polarized
vacuum around a positive charge $+Q$.} \label{fig.1}
\end{figure}

The intensity of vacuum polarization can be defined as
$\mathbf{P'}=\sum\mathbf{p'}/V$, where $\mathbf{p'}$ is the
electrical dipole moment of a single polarized virtual charge pair,
and $V$ is a volume. If there are $N$ virtual vacuum dipoles in
volume $V$, each dipole has an electrical dipole moment
$\mathbf{p'}=q'\cdot 2\mathbf{a}$ ($2a$ is the average distance
between the two vacuum virtual charges $\pm q'$ --- the reason we
say ``average'' here is that, for an individual virtual dipole, the
length $2a$ can be quite different due to the uncertainties in
quantum mechanics), we have
\begin{equation}
\mathbf{P'}=\frac{N}{V}q'\cdot 2\mathbf{a}.
\end{equation}

Eq.~(2) can be rewritten as $\mathbf{E}=\mathbf{P'}/\varepsilon_0 $,
which means that an electric field is just a polarized distribution
of the vacuum virtual charge pairs. At a given point, the field
direction depends on the polarization direction. And the field
intensity is directly proportional to the intensity of vacuum
polarization or to the areal density of virtual polarization charges
of the vacuum. For the vacuum around a real charge $Q$, the areal
density of virtual polarization charges $\sigma'$ at a distance $r$
from the centre of the charge $Q$ is $\sigma'=Q'/4\pi r^2$, where
$Q'$, being equal to $Q$ in magnitude, is the total virtual
polarization charges of the vacuum on the spherical Gaussian surface
of radius $r$. Therefore, the field intensity is found to be
\begin{equation}
E=\frac{P'}{\varepsilon_0}=\frac{\sigma'}{\varepsilon_0}=\frac{1}{4\pi
\varepsilon_0}\frac{Q}{r^2}.
\end{equation}

Under the above interpretation of electric field, the field energy
in a vacuum can be understood as the polarization energy of the
vacuum, i.e., the potential energy of the virtual polarization
charges. That is, the energy density of an electric field (EF) in
vacuum will be $w_\textnormal{{EF}}=N(2 \cdot \frac{1}{2} k
a^2)/2aS$, where $N$ is the number of virtual vacuum dipoles in a
volume of $V=2aS$, $S$ is an area perpendicular to the direction of
polarization, and the areal density of virtual polarization charges
is $\sigma'=Nq'/S$, $k$ is the stiffness factor, which satisfies
$F=ka=q'E=q'\sigma'/\varepsilon_0$. Combining these relations, we
have
\begin{equation}
w_\textnormal{{EF}}=\frac{1}{2}\frac{\sigma'^2}{\varepsilon_0}=\frac{1}{2}\varepsilon_0
E^2.
\end{equation}

Using the concept of vacuum polarization, the force exerted by an
electric field on a test charge can be analyzed as the action of the
nearby virtual dipoles in vacuum. Fig.~2 sketches the vacuum
polarized by the resultant field of an external field $\mathbf{E}$
and a test charge $+q$, where we can see that there will be an
attraction between the test charge and the nearby virtual dipoles.
But the vacuum on the right side of the test charge is more
polarized than that on the left side. Then the test charge tends to
move to the right side, i.e., in the direction of the external
field. This is what we say that the test charge is exerted an
electrical force $\mathbf{F}=q\mathbf{E}$.

\begin{figure}
\includegraphics[width=2.6in]{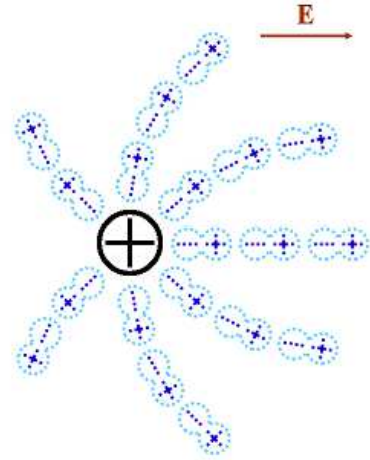} \caption{Vacuum
polarized by the resultant field of an external field \textbf{E} and
a test charge $+q$.} \label{fig.2}
\end{figure}

Second, examining the magnetic intensity vector \textbf{H} in a
magnetic medium
\begin{equation}
(-\mathbf{H})=\mathbf{M}+\left(-\frac{\mathbf{B}}{\mu_0}\right),
\end{equation}
where $\mathbf{M}$ is the magnetization intensity of the medium,
$\mu_0$ is the permeability of vacuum, \textbf{B} is the magnetic
induction intensity, the last term $(-\mathbf{B}/\mu_0)$ can be
conceived as the magnetization of vacuum $\mathbf{M}'$, that is
\begin{equation}
\left(-\frac{\mathbf{B}}{\mu_0}\right)=\mathbf{M'}=\frac{\sum\mathbf{\Omega'}}{V}=\frac{N}{V}i'S\mathbf{s}_0,
\end{equation}
where $\mathbf{\Omega'}=i'S\mathbf{s}_0$ is the magnetic moment of a
vacuum virtual circular current $i'$ with an enclosed area $S$,
$\mathbf{s}_0$ is the unit vector, being opposite to the magnetic
direction, $N$ is the total number of the virtual circular currents
in volume $V$.

\begin{figure}
\includegraphics[width=2.6in]{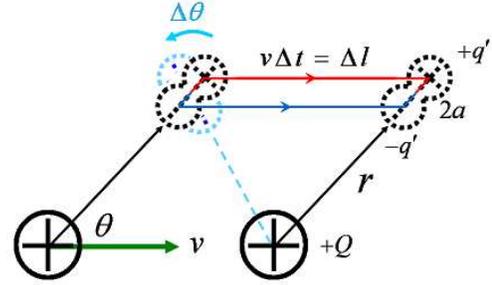} \caption{The magnetization of
vacuum around a right moving charge $+Q$.} \label{fig.3}
\end{figure}

Generally, there is neither free charge nor molecular current in a
vacuum. But still, there could be virtual circular currents formed
in a magnetized vacuum. This is illuminated in Fig.~3, where a
charge $+Q$ is moving rightward in vacuum. From the figure, we can
see that the vacuum polarization will be in a continuous
rearrangement when the charge $Q$ is in a directional movement. For
example, the virtual dipole at a certain point in vacuum will rotate
an angle of $\Delta \theta$ when $Q$ moves a distance of $\Delta l=v
\Delta t$ (here $v$ is the velocity of $Q$, $\Delta t$ is the time
interval). At the same time, the original direction of polarization
at this point will translate a distance of $\Delta l$. This
translation is equivalent to a virtual circular current, which is
lined out with a parallelogram in the figure (where the right
translation of $-q'$ equals to a left translation of $+q'$). It is
the virtual currents of this type that constructed the magnetic
field \textbf{B} in vacuum. Using Eqs.~(2), (3) and (7), the
magnetic field in Fig.~3 can be deduced as
\begin{eqnarray}
\mathbf{B} &=& -\mu_0\frac{N}{V}\frac{q'}{\Delta t}2a\Delta l \sin
\theta \mathbf{s}_0  \nonumber \\
 &=& -\mu_0 v P'\sin \theta
\mathbf{s}_0 \nonumber\\
 &=& -\mu_0\varepsilon_0 E v \sin \theta
\mathbf{s}_0 \nonumber\\
 &=& \frac{\mathbf{v}}{c^2} \times \mathbf{E},
\end{eqnarray}
where $c=1/\sqrt{\varepsilon_0 \mu_0}$ is the velocity of light in
vacuum.

The above result is just in agreement with what we have known in
electromagnetics and is correct in the cases of relativity. It
indicates that a magnetic field is also based on the polarization of
vacuum, or in more detail, a magnetic field is a rearrangement of
the vacuum polarization which contains a series of virtual circular
currents in vacuum. These currents are in fact a visual effect of
the rotation of the vacuum virtual dipoles caused by the movement of
charge $Q$. This rotation as shown with a curved arrow and $\Delta
\theta$ in Fig.~3 corresponds to a kinetic energy of the vacuum
virtual charges, which is just the energy of a magnetic field.
Considering that the angular velocity of the dipole rotation is
$\omega=d\theta /dt=v \sin\theta /r$, the energy density of the
magnetic field (MF) in vacuum around a moving charge $Q$ can be
given as
\begin{equation}
w_\textnormal{{MF}}=\frac{N}{V}\cdot 2\cdot \frac{1}{2}m'
(a\omega)^2=\frac{1}{2}\frac{B^2}{\mu_0}\cdot
\left(\frac{4\pi}{\mu_0}\cdot\frac{m'}{q'}\cdot\frac{a}{Q}\right),
\end{equation}
where $m'$ is the equivalent mass of the virtual charge $q'$. This
equation is just the well-known formula of magnetic energy density
$w_\textnormal{{MF}}=B^2/ 2 \mu_0$ when
\begin{equation}
\left(\frac{4\pi}{\mu_0}\cdot\frac{m'}{q'}\cdot\frac{a}{Q}\right)=1,
\end{equation}
where $m'/q'$ is the mass-to-charge ratio of a vacuum virtual
charge, and $a/Q$ is the ratio of the dipole average half length to
the moving charge $Q$.

Similar to the electrical force, we say that the magnetic force
exerted on a test moving charge can be understood as the action of
the nearby virtual circular currents in the magnetized vacuum.
Fig.~4 sketches such an interaction, where the magnetization
intensity of vacuum caused by an external uniform magnetic field
\textbf{B} and a moving charge $+q$ is represented by the chromatic
diagram and the contour lines. In each contour map, the line labeled
by ``0'' denotes the boundary between the two opposite (i.e., the
clockwise and anticlockwise) circular currents in vacuum. According
to the figure and what we have known about the behavior of a moving
charge in an external magnetic field, it is reasonable to suppose
that a moving charge tends to move along this boundary. Since the
larger the external field \textbf{B}, the more curved the boundary
line as shown with thick arrows in the figure, we then know that the
stronger the field \textbf{B}, the heavier the force \textbf{F}
exerted perpendicularly on the moving charge. It is compatible with
the definition formula $\mathbf{F}=q\mathbf{v} \times \mathbf{B}$.\\

\begin{figure}
\includegraphics[width=2.6in]{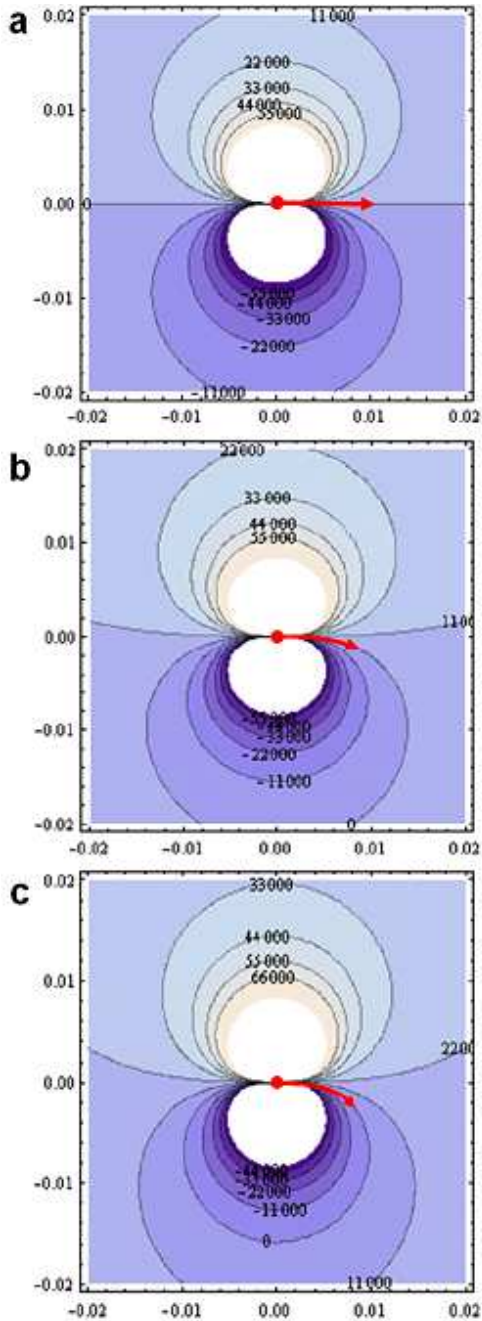} \caption{The vacuum magnetized by
an external uniform magnetic field \textbf{B} pointing out of the
page and a right moving charge $+q$ simultaneously: (a) $B=0$; (b)
$B=B_0$; (c) $B=2B_0$.} \label{fig.4}
\end{figure}

The above analysis of the electromagnetic energy and force
demonstrated that there can be an intrinsic connection between the
electromagnetic field and the quantum vacuum. That is to say, both
the electric field and magnetic field can be understood as the
effect of vacuum polarization. To emphasize this, we will present
two key points below: \ \\ \

(1) \emph {An electric field is a polarized distribution of the
vacuum virtual dipoles, and a magnetic field in vacuum is a
rearrangement
 of the vacuum polarization.}\\

This result can help us understand the electromagnetic phenomena on
a more profound basis. For example, the formation of electromagnetic
wave can then be interpreted through the vacuum polarization. Since
an electric field is a polarized distribution of the vacuum virtual
dipoles, when an electric field is changed, there will be a virtual
polarization current occurring in the vacuum. This polarization
current, or the so-called ``displacement current'', will then cause
a magnetic field. On the other hand, a magnetic field is a
rearrangement of vacuum polarization, or in other words, a rotation
of the virtual dipoles, which in effect forms the virtual circular
currents in vacuum. When a magnetic field is changed, there will be
a tendency of keeping the original dipole rotations or the original
vacuum circular currents, which serves as a rotational electric
field. In brief, a changing electric field generates a magnetic
field, and a changing magnetic field induces an electric field. The
repetition of this process gives rise to a propagation of
electromagnetic wave in space.\ \\ \

(2) \emph {The vacuum virtual dipoles around an elementary charge
$e$ have a characteristic half length on average:}
\begin{equation}
a=\frac{\mu_0}{4\pi}\cdot\frac{q'}{m'}\cdot Q,
\end{equation}
which is obtained from Eq.~(10). Substituting the charge-to-mass
ratio $q'/m'=e/m_\textnormal{e}=1.758 \times
10^{11}\textnormal{C/kg}$, and the elementary charge $Q=e=1.6 \times
10^{-19} \textnormal{C}$ into Eq.~(11), we found that:
\begin{equation}
a=2.8 \times 10^{-15} \textnormal{m}.
\end{equation}
It is interesting to find that this length is just the electron
classical radius or about the size of a proton.\\

This result will lead to further findings. For example:\\

a) The step distribution of the energy density of the electric field
around an electron.\\

According to Eqs.~(3), (4), (5) and (11), we see that there can be a
step change of the energy density of the electric field around an
electron as shown in Fig.~5, where the step width $2a$ is the
average length of the vacuum virtual dipoles, and the energy density
unit is $A=e^2/32\pi^2\varepsilon_0a^4$; whereas in the classical
theory the energy density of the field is inversely proportional to
$r^4$ in a smooth way ($r\in[0,\infty]$ is the distance from the
electron).

Just as that the energy quantum solved the problem of ``ultraviolet
catastrophe'' in black body radiation, the step distribution of
energy density will eliminate the divergence in calculating the
electrostatic energy of an electron. This divergence, as we know
that, is inevitable in the classical theory of electromagnetism. In
the quantum theory of fields, the divergence is constrainedly evaded
through the introduction of renormalization. But here, according to
the step distribution of the energy density as shown in Fig.~5, the
calculation of the electron's electrostatic energy turns to be:
\begin{eqnarray}
W &=& \int w_\textnormal{{EF}}dV  \nonumber \\
 &=& \int \frac{1}{2}\varepsilon_0 E^2 \cdot
4 \pi r^2dr \nonumber\\
 &\sim& \int_{a/2}^\infty \frac{1}{2}\varepsilon_0 \left(\frac{1}{4\pi
\varepsilon_0}\frac{e}{r^2}\right)^2 \cdot
4 \pi r^2dr  \nonumber\\
 &=& \frac{1}{4\pi \varepsilon_0}\frac{e^2}{a}=m_\textnormal{e} c^2\neq
 \infty.
 \end{eqnarray}
This calculation differs from the semiclassical models incorporating
short-distance cutoffs as the work of Boyer \cite{Boyer1969}, etc.,
both in method and in idea.\\

 \begin{figure}
\includegraphics[width=3.0in]{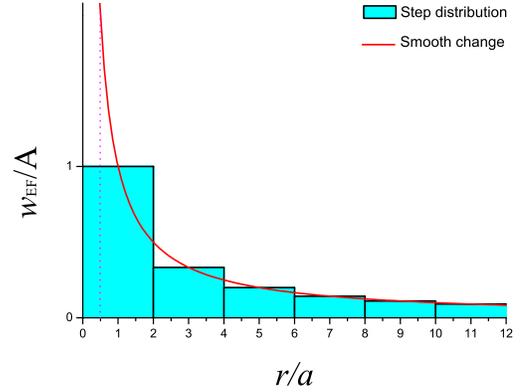} \caption{The energy density
distribution of the electric field around an electron.}
\label{fig.5}
\end{figure}

 b) The relation between the fine structure constant and
the vacuum polarization distribution.\\

Ever since it was discovered, the fine structure constant
$\alpha=e^2/4\pi \varepsilon_0 \hbar c$ has been a great mystery.
Now the step distribution of the vacuum polarization energy offers
us a clue to understand the constant. From Eq.~(11) we know that,
for a moving electron in an atom, the corresponding average half
length $a$ of the surrounding virtual dipoles satisfies
\begin{equation}
\frac{a}{R}= \alpha^2,
\end{equation}
where $R$ is the Bohr radius. This equation implies that there
exists a deep relation between the fine structure constant and the
vacuum polarization.

With the above relation, we can then calculate the energy difference
when the electron in a ground-state hydrogen atom has a change of
radial position $\Delta r=2a$ from the Bohr radius $R$, that is
\begin{equation}
\Delta W = \left| \Delta \left(-\frac{1}{8\pi
\varepsilon_0}\frac{e^2}{r}\right)\right|= \left| -\frac{1}{4\pi
\varepsilon_0}\frac{e^2}{R}\right|\alpha^2,
 \end{equation}
which is just the energy order of the atomic fine structure.\\

c) An extremely high energy density of the electromagnetic field.\\

Fig.~5 indicates that there can be an extremely high energy density
of the electric field around an electron at $r\in[0,2a]$, that is:
\begin{eqnarray}
w_\textnormal{{max}} &\sim & \frac{1}{2}\varepsilon_0 E^2_a \nonumber\\
 &=& \frac{1}{2}\varepsilon_0 \left(\frac{1}{4\pi
\varepsilon_0}\frac{e}{a^2}\right)^2  \nonumber\\
 &=& 1.5 \times 10^{29} \textnormal{J/m}^3.
 \end{eqnarray}
The order of this result can also be obtained supposing that all the
virtual dipoles are composed of charge pairs $e^--e^+$ with the
average space between every two neighboring dipoles being $2a=5.6
\times 10^{-15}$m.

Since the energy density of a magnetic field has the same order as
that of an electric field, the above value can be considered as a
corresponding extremely high energy density of the electromagnetic
field. This energy density can be converted into an optical power
density:
\begin{eqnarray}
I= w_\textnormal{{max}} \cdot c  \sim 10^{33} \textnormal{W/cm}^2,
 \end{eqnarray}
which is 4 order higher than that of the Schwinger critical
intensity ($\sim$$10^{29} \textnormal{W/cm}^2$) for $e^--e^+$ pair
production in vacuum \cite{Narozhny2004}.

The above obtained extremely high energy density of the
electromagnetic field or extremely high optical power density is
significant to the ultrastrong-field physics, the pursuing of super
powerful light source and the pair production in vacuum,
etc.\\

In summary, according to the analysis of the energy and force of the
electric and magnetic fields on the basis of vacuum polarization, it
is concluded that an electric field is a polarized distribution of
the vacuum virtual dipoles, and that a magnetic field in vacuum is a
rearrangement of the vacuum polarization. Thus, the electromagnetic
wave can be regarded as a successional changing of the vacuum
polarization in space. Also, it is found that the virtual dipoles
around an elementary charge possess an average half length $a=2.8
\times 10^{-15}$m. This result leads to the knowledge that an
electric field has a step distribution of the energy density, which
eliminated the divergence in calculating the electron's
electrostatic energy. And it is known that there is a relation
between the fine structure constant and the vacuum polarization
distribution, which reduced the mystery of the constant $\alpha$.
Finally, it is figured out that an extremely high energy density of
the electromagnetic field can be $\sim10^{29} \textnormal{J/m}^3$,
which implies an optical power density $\sim10^{33}
\textnormal{W/cm}^2$ far higher than the Schwinger critical value.
With these interesting findings, we anticipate that the vacuum
polarization investigation of the fields will be developed further
and applied to more fundamental problems of physics.

This work was supported by Hangzhou Dianzi University (grant no.
KYS075608069).

\ \


\


\appendix{\textbf{}}

\end{document}